\documentclass[conference]{IEEEtran}
\IEEEoverridecommandlockouts

\makeatletter
\def\markboth#1#2{\def\leftmark{\@IEEEcompsoconly{\sffamily}\MakeUppercase{\protect#1}}%
\def\rightmark{\@IEEEcompsoconly{\sffamily}\MakeUppercase{\protect#2}}}
\makeatother

\usepackage[english]{babel}
\usepackage{xcolor}
\usepackage{lipsum}
\usepackage{caption}
\usepackage{cite}
\usepackage[pdftex]{graphicx}
\usepackage{subcaption}
\usepackage{amsmath}
\usepackage{amsfonts}
\usepackage{amssymb}
\usepackage{amsthm}
\usepackage{array}
\usepackage{verbatim}
\usepackage{listings}
\usepackage{algorithm}
\usepackage{algpseudocode}
\usepackage{url}
\usepackage{enumerate}
\usepackage{multirow}

\usepackage{epsfig}
\usepackage{epstopdf}
\usepackage{multicol}
\usepackage[font=footnotesize]{caption}

\newcommand{\bi}{\begin{itemize}}
\newcommand{\ei}{\end{itemize}}
\newcommand{\be}{\begin{equation}}
\newcommand{\ee}{\end{equation}}

\def\beq{\begin{equation}}
\def\eeq{\end{equation}}
\def\beqa{\begin{eqnarray}}
\def\eeqa{\end{eqnarray}}
\def\beqan{\begin{eqnarray*}}
\def\eeqan{\end{eqnarray*}}

\title{The Potential of Resource Sharing\\ in 5G Millimeter-Wave Bands}
\author{{{\bf Mattia Rebato}$^\dagger$, {\bf Marco Mezzavilla}$^\diamond$, {\bf Sundeep Rangan}$^\diamond$, {\bf Michele Zorzi}$^\dagger$ }\\
$^\diamond$NYU Tandon School of Engineering, Brooklyn, NY \\
$^\dagger$ University of Padova, Italy
}

\begin{document}
\maketitle

%


\begin{abstract}
With the severe spectrum shortage in conventional cellular bands, the millimeter
(mmWave) frequencies, roughly above 10~GHz, have been attracting growing attention 
for next-generation micro- and pico- cellular wireless networks.
A fundamental and open question is how these bands should be used by cellular operators.
Cellular spectrum has been traditionally allocated following an exclusive ownership model.
However, in this paper we argue that the distinct nature of mmWave communication 
-- the massive bandwidth degrees of freedom, directional isolation and
high susceptibility to blockage -- suggest that spectrum and infrastructure sharing
between multiple operators may be necessary to exploit the full potential of these bands.
High-level capacity analyses are presented that reveal
significant possible gains under spectrum and infrastructure sharing, even under minimal 
coordination between operators.  Moreover, we discuss how network technologies including 
software defined networks (SDNs) and network function virtualization (NFV) 
can easily enable resource sharing 
by having a programmable core entity provide 
transparent inter-operator access to the end user.
\end{abstract}

\section{Introduction}

The millimeter wave (mmWave) frequencies, roughly between 30 and 300 GHz,
 \footnote{While the millimeter-wave spectrum is defined as the band between 30-300 GHz, industry has loosely considered mmWave to be any frequency above 10 GHz.} are a new and promising frontier for cellular wireless communications~\cite{RanRapE:14}. 
With the rapidly growing demand for cellular data rate, conventional frequencies below 3 GHz are now highly congested. For example, in the most recent FCC auction, 
65 MHz of AWS-3 spectrum sold for a record breaking \$45 billion, underlining this severe spectrum crunch in expanding networks today. In contrast, the mmWave bands over vast, largely untapped spectrum – up to 200 times all cellular allocations by some estimates.  
Due to this enormous potential, mmWave networks have been widely cited as one of the most promising technologies for Beyond 4G and 5G cellular evolution.

A basic question for the development of mmWave systems is how these new bands
should be allocated to and used by cellular carriers.
Cellular spectrum in conventional bands
has been traditionally allocated in a long-term license model
where each carrier has exclusive use of the frequencies.  This model significantly
simplifies design
by eliminating uncontrolled interference and has been the basic design
assumption in virtually every cellular standard.  
Cellular carriers also often own and manage
infrastructure and core network elements and have exclusive usage (through ownership or
leasing) of the wired backhaul.

However, with the introduction of mmWave bands, this
highly-vertically integrated model for cellular networks may
no longer provide an efficient use of the spectrum or infrastructure resources. 
Specifically, in the mmWave space, as we will see in simulations below,
the massive bandwidth and
spatial degrees of freedom are unlikely to be fully used by any one cellular operator. 
In addition, due
to the intermittent and highly localized nature of traffic and channel capacity, backhaul and wired
networks resources will experience high levels of variability. This extremely bursty demand profile necessitates some form of statistical multiplexing in the backhaul and wired portions of the network to provision network capacity appropriately. Furthermore, mmWave signals experience high penetration loss through brick and glass, and a dense indoor infrastructure is needed for coverage.
Deployment of such infrastructure separately by each operator could be prohibitively 
costly, so sharing of
base stations (BS) and relays may provide a cost effective 
way to obtain the required density.

Regulatory authorities such as the United States
Federal Communications Commission (FCC) have already recognized these issues.
For example, in their recent
notice of inquiry (NoI)\footnote{Federal Communications Commission document: ?In the Matter of Use of Spec- trum Bands Above 24 GHz For Mobile Radio Services,? Notice of Inquiry in GN Docket No. 14-177.}, among other insightful considerations, promotes sharing-related research directions within the mmWave bands: \emph{The technologies being developed for advanced mobile applications in frequencies above 24 GHz could allow opportunities for reuse of spectrum and for spectrum sharing that are not possible at lower frequencies with current technology.}

%


\subsection{Related work}
\label{rel_work}

Resource sharing has common challenges with heterogeneous networks (HetNets). Although densification has observable limits for microwave frequencies, it is shown in \cite{Baldemair:15} that denser deployments are advantageous for mmWave bands because of the different propagation characteristics for NLOS and LOS environments.

\textbf{Spectrum sharing:} In the \emph{microwave bands}, interference is a limiting factor. Competitive and greedy sharing methods might cause underutilization of the spectrum, as shown in \cite{Anchora:11}. A viable sensing approach for dynamic inter-operator spectrum sharing for an LTE-A system with carrier aggregation (CA) is proposed in \cite{Gerstacker:14}. Additionally, under the assumption of partial interference suppression, the optimality of full spectrum sharing is validated by means of simulations in \cite{Zorzi:13}.


In the \emph{mmWave bands}, spectrum sharing results are mixed. Under dense deployments, interference avoidance gives optimum results for WiGig \cite{Zeng:14}. However, it is shown that directional transmission allows considerable throughput gain even with blind reuse of frequency bands \cite{Shi:14}, where a ray-tracing model is used to characterize the channel. The authors in \cite{Schulz:14} propose an interference sensing beamforming mechanism in wireless personal access networks (WPANs) that outperforms blind selection algorithms by $15\%-31\%$. In \cite{texas}, the authors show that sharing spectrum licenses increases the per-user rate when antennas have narrow beams; additionally, all networks can share licenses with less bandwidth and still achieve the same per-user median rate as if each had an exclusive license with more bandwidth.

\textbf{Infrastructure sharing:} Load-aware strategies targeting infrastructure sharing for \emph{microwave bands} are proposed in \cite{ElSawy:13}: the authors propose an approach based on cognition capabilities, which are achieved through spectrum sensing. In \cite{infra-sha}, the authors investigate the current technological, regulatory, and business landscape from the perspective of sharing network resources, and propose several different approaches and technical solutions for network sharing.

In \cite{texas}, BS co-location in the \emph{mmWave bands} is considered. If compared to our framework, the multi-antenna and propagation models adopted by the authors are simplified, for the sake of the analysis; nonetheless, the results shown can still capture promising performance trends.

\subsection{Contribution of this article}
\label{contribution}

The broad purpose of this article is to provide some assessments of how
much gain spectrum and infrastructure sharing can obtain, and how these gains can be achieved.  Toward this end, we provide the following:
\begin{itemize}
\item \emph{Scaling laws:}  To provide high level insights, we 
first derive simple
scaling laws on how we would expect capacity to scale with density of cells
with and without sharing.  We identify three possible regimes -- interference-limited,
power-limited and outage-limited.  
Cellular networks in current frequencies are interference
limited where capacity scales roughly linearly with base station density.  However,
since mmWave links will have very wide bandwidths and
are highly susceptible to blockage, they will likely reside in power and outage-limited
regimes instead.  These regimes can  result in \emph{super-linear} scaling of capacity with density,
suggesting a fundamentally better capacity scaling with sharing.  We confirm these
results via some simple simulations using measurement based channel models
at both conventional and mmWave frequencies.

\item \emph{Capacity evaluation:} Using a standard 3GPP evaluation 
methodology\footnote{Evolved universal terrestrial radio access (E-UTRA); radio frequency (RF) system scenarios, TR 36.942 (Release number 12).} combined with realistic 
measurement-based channel models,
we provide high-level capacity estimates under different sharing scenarios.
Our results confirm that significant gains are possible when spectrum and access
infrastructure can be shared
by multiple operators.
Importantly, 
our preliminary results reflect scenarios where resources follow a \emph{blind}, \emph{uncoordinated} allocation scheme in the radio access network (RAN), 
and show that this simplified approach actually performs very close to a fully-coordinated scheme.

\item \emph{Virtualized architecture}: 
Due to advances in software defined networks (SDNs) and 
network function virtualization (NFV),
the forthcoming 5G architecture is foreseen to be largely virtualized. 
This architecture can facilitate spectrum and infrastructure sharing 
 by having a flexible (programmable) core entity 
 enable transparent inter-operator access to the end user.
\end{itemize}

The rest of the article is organized as follows.
In Section \ref{densif}, we discuss the importance of network densification in mmWave bands. In Section \ref{potential}, we describe various sharing configurations, which include \emph{access}, \emph{infrastructure} and \emph{spectrum}. A preliminary numerical evaluation, along with some important remarks, is reported in Section \ref{use}. Finally, we conclude our article and describe some future research directions in Section \ref{conclusion}.

\section{Capacity and Density Scaling}
\label{densif}

Before conducting a more detailed analysis, we can gain insights
into the value of spectrum and infrastructure 
sharing in the mmWave bands by looking at some simple scaling laws.
Specifically, we will argue the following:  (i) sharing infrastructure and
spectrum amongst $M$ operators increases the effective density of cells 
available to each UE 
by a factor of $M$; (ii) under some simple models,
increasing density has significant benefits for mmWave
systems with many degrees of freedom and subject to outage;
and (iii) these simple calculations be validated with more detailed simulations
under realistic channel models.

\paragraph*{Densification via sharing}

We first argue that sharing amongst $M$ operators increases the effective 
density of cells and the available bandwidth by a factor of $M$.
To see this, 
suppose there is a total bandwidth $W$ that has to be allocated amongst $M$ networks
and each network has a total of $N_{UE}$ mobiles and $N_{BS}$ base station cells.
Now, consider two possible sharing scenarios: 
(i) \emph{no sharing},  where each network
has a bandwidth $W/M$ and the UEs in the network can only connect 
to the base station cells of that network; and 
(ii) \emph{complete sharing}, where all $MN_{UE}$ mobiles can access all $MN_{BS}$
cells across the entire bandwidth $W$.

Now, for both sharing and no sharing, there are $N_{UE}/N_{BS}$
UEs per cell.  Hence, if we let $W_{UE}$ be the available bandwidth 
per UE, with sharing $W_{UE}=WN_{BS}/N_{UE}$ and without sharing
$W_{UE}=WN_{BS}/(MN_{UE})$.  Hence, sharing increases the available bandwidth
per UE by a factor of $M$.  Also, with sharing, each UE can access $M$ times the 
number of BS cells, so it sees an \emph{effective} density increase by $M$.

\begin{figure}[t!]
    \includegraphics[width=\columnwidth]{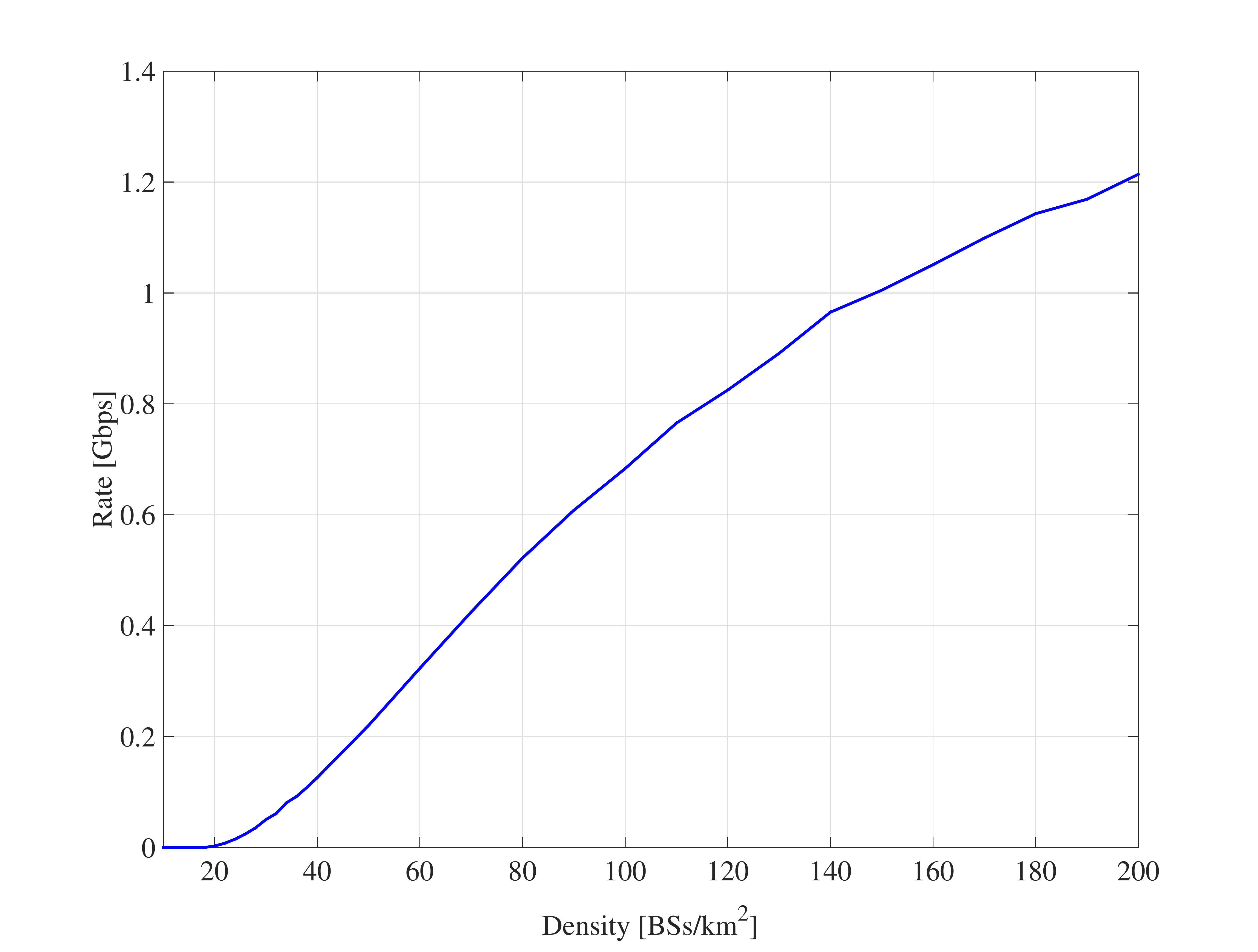}
      \caption{Actual scaling of the 5\% user rate as a function of density.}
      \label{5perc}
\end{figure}

\begin{figure}[t!]
    \includegraphics[width=\columnwidth]{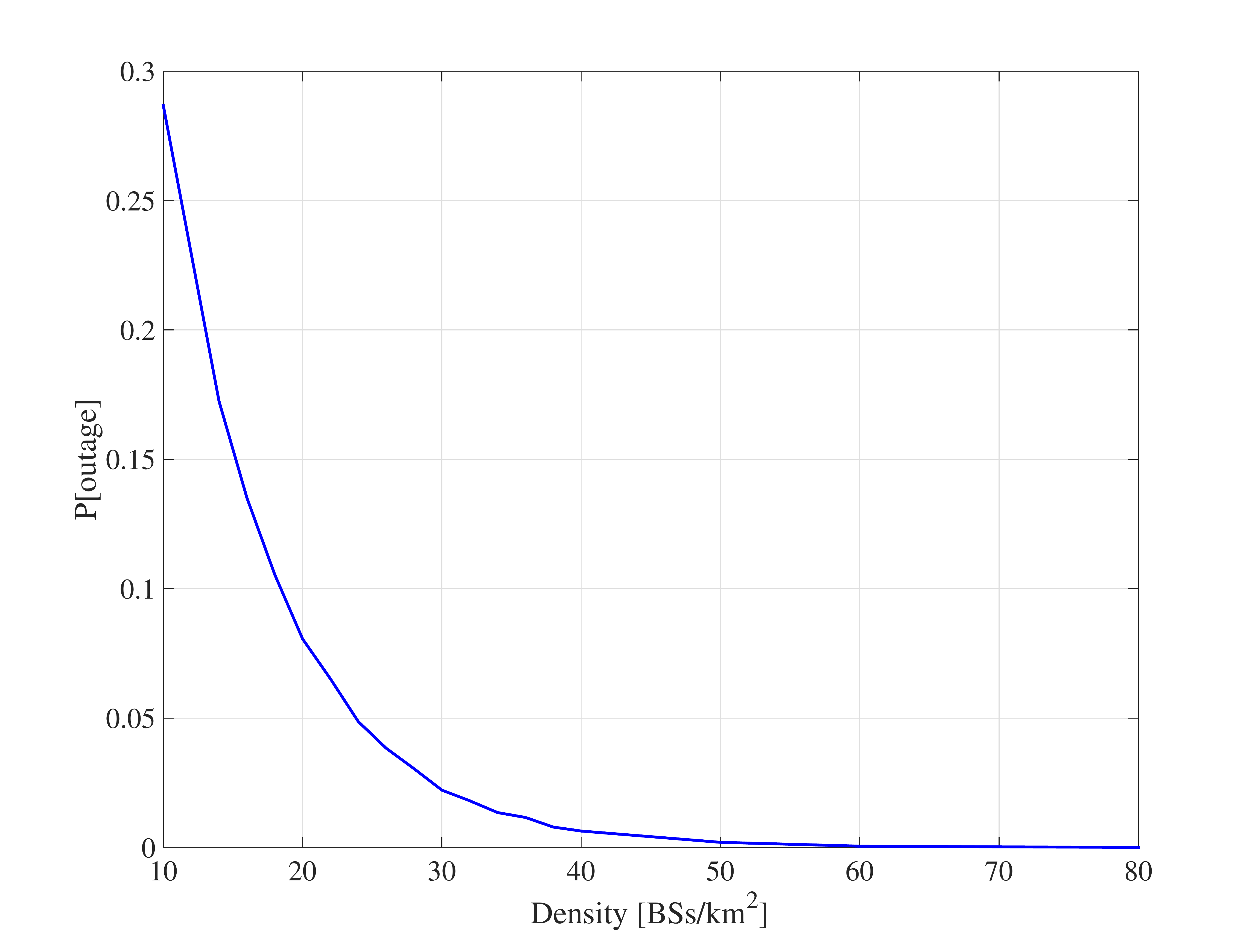}
      \caption{Decrease in outage probability with cell density.}
      \label{outage_desity}
\end{figure}

\paragraph*{Improvements with density}
Given that sharing effectively increases the cell density, let us look at how
capacity scales with density.  
To this end, 
let $\rho$ be the density of base stations (base stations per square kilometer)
so that $N_{BS} = A\rho$ where $A$ is the total area. 
Fix the number of UEs, $N_{UE}$, bandwidth $W$, 
and network area, $A$, and consider scaling $N_{BS}$ with 
the density $\rho$.  We look at two key quantities:  the available bandwidth
per UE, $W_{UE}$, and the average distance to the closest cell, which 
we will denote by $d_{UE}$.

From the above calculations, the available 
bandwidth per UE is $W_{UE} \propto N_{BS} \propto \rho$, i.e., it grows linearly with $\rho$.
We will write this as $W_{UE} = O(\rho)$.  Also, since the cell area
scales as $O(1/\rho)$, the distance to the closest base station cell 
should scale as $d_{UE} = O(\rho^{-1/2})$.  If we assume that the base station
transmit power is fixed and all cells transmit across the entire
bandwidth, then the received power should scale as $P_{UE} = O(d_{UE}^{-\alpha})$,
where $\alpha$ is the path loss exponent.  Equivalently, $P_{UE} = O(\rho^{\alpha/2})$.

With these preliminary computations, we can estimate the scaling of the rate per UE, $R_{UE}$,
as a function of the density, $\rho$.  First, suppose that the network is 
interference-limited, meaning that the received power over the bandwidth is much 
higher than the thermal noise density.  This is the case in virtually all micro-cellular
networks that have very limited bandwidths with cells that are close.  
Increasing cell density brings serving base stations closer to users,
but it also brings interfering base stations closer.  If the received
signal and interference powers scale by the same factor $d^{-\alpha}$,
and the interference dominates the thermal noise, the signal-to-interference
and noise 
ratio (SINR) per user does
not increase with density. Hence, in the interference-limited regime,
the capacity per unit bandwidth will be the same,
and thus the rate per UE, $R_{UE}$, scales with the bandwidth per UE, $W_{UE}$.
From the above calculations, we find $R_{UE}=O(\rho)$.

The unique feature of mmWave systems 
is that due to wide bandwidths and large number of antennas, 
the number of degrees of freedom per UE can be large.  Hence, even in micro-cellular deployments
with relatively small cells, transmissions may be
power-limited instead of bandwidth-limited.  In this regime, the rate will scale linearly
with power.  From the above calculations $R_{UE} = O(\rho^{\alpha/2})$.
In non-line-of-sight (NLOS) regimes, $\alpha > 2$, and rate scaling is \emph{super-linear} with density.
We thus see that since many links in a mmWave system will be 
NLOS and power-limited, increasing density -- and hence enabling sharing -- can 
provide fundamentally greater improvements in rate than in current interference-limited
systems.

MmWave systems will also benefit in a second manner from increased density,
not possible at current frequencies.  MmWave signals are highly susceptible to 
blockage from many building materials 
and thus many locations may be in outage.  Outage in this manner occurs much
less frequently in current frequencies.  Increasing cell density can overcome
this outage by increasing the chance that there is a cell within range.
To estimate how much the outage probability will decrease with
cell density, suppose, as a very simple model, that each cell can cover a fixed
area of $A_c$ and that, outside this area, the signal from the base station is blocked.
Now, if the density of cells is $\rho$, each cell must cover a total area
of $1/\rho$ and hence a fraction $\max\{1-A_c\rho,0\}$ of UEs will be in outage.
Thus, under this simple model, the outage probability will roughly decrease to zero
linearly with density.

\paragraph*{Scaling under realistic models}  
To validate the above simple scaling laws under realistic models, 
the left panel of Fig.~\ref{5perc} shows the actual scaling of the median and 
5\% user rate as a function of density.  Here, we have followed the 
simulation methodology in \cite{mustafa} which uses mmWave channel models
derived from actual measurements at 28~GHz in New York City \cite{mustafa}. Details of the simulation,
such as noise power, transmit power, beamforming, etc. can all be found in \cite{mustafa}.
Statistical channel models in those works derive a path loss exponent of $\alpha \approx 2.7$ and thus predict super linear rate increases at low density.
The left panel of Fig.~\ref{5perc} shows exactly this predicted behavior.
At lower densities, the capacity clearly scales greater than linearly and only at very high
densities of 80 BSs/km$^2$ (corresponding to cell radii of 63 m), does the
capacity start to scale linearly.  Note that at very high densities, the scaling
is slightly less than linear.  This is an artifact of the simulation such that
at very high densities, there is often less than one UE per cell, which is not a
reasonable regime.  
\begin{figure*}[t!]
    \includegraphics[width=\textwidth]{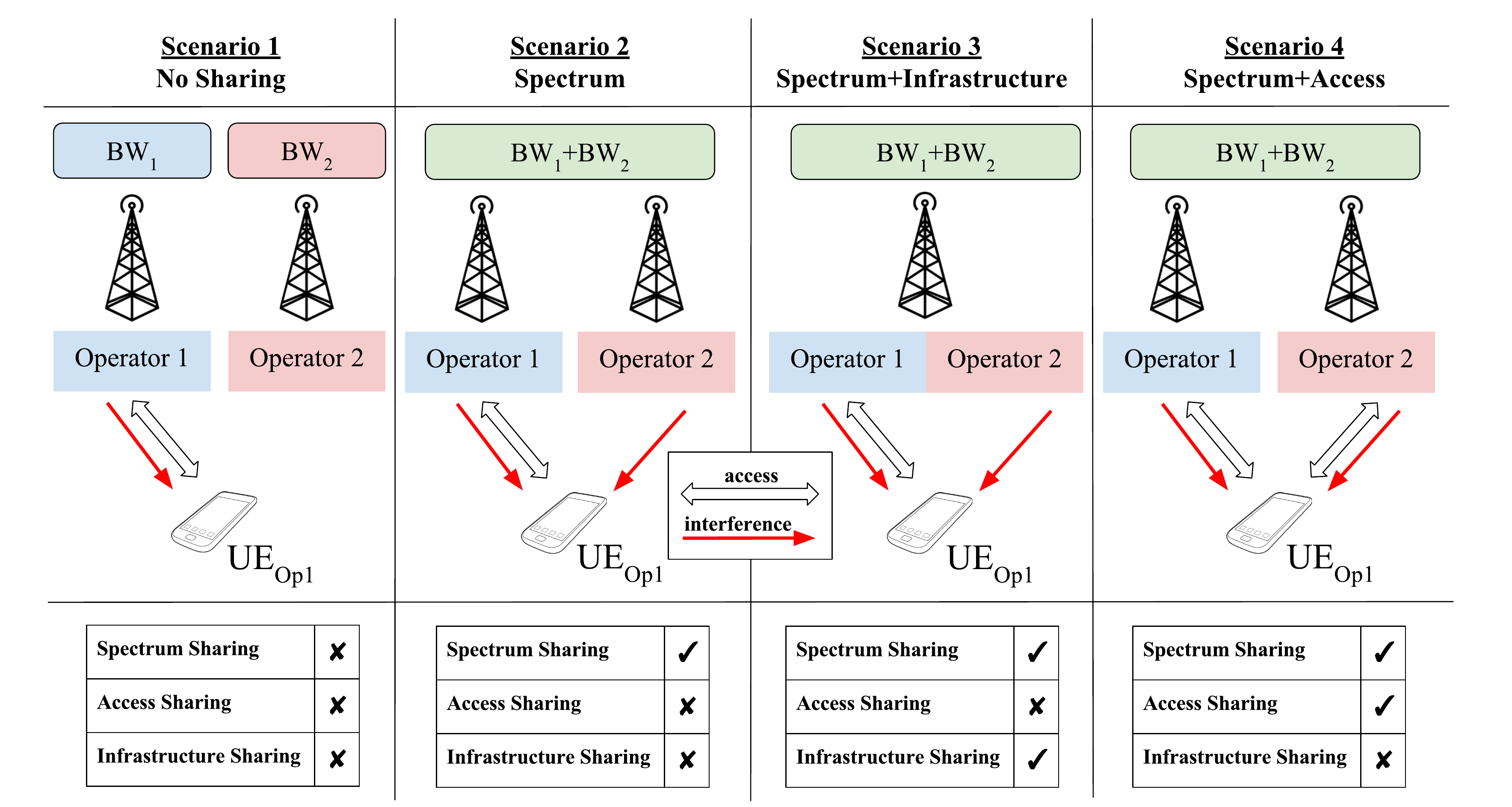}
      \caption{Graphical representation of the sharing paradigms considered.}
      \label{shar_conf}
\end{figure*}

The right panel of Fig.~\ref{outage_desity} shows the decrease in outage with cell density.
Here, we have defined outage as the fraction of users that could not obtain a
minimum target rate.  As predicted, there is a sharp decrease in this rate with cell
density.  

We conclude that the spectrum and infrastructure sharing, 
which has the effect of increasing cell density, has two major benefits for
mmWave systems, not available at current frequencies.  First, there is a super-linear
increase in rate due to systems being power-limited; and second, there is 
a decrease in outage probability due to the signals being blocked.

\section{Sharing Paradigms}
\label{potential}

Capturing some preliminary insights over this emergent sharing-based technology is critical to better drive the business opportunity. To do so, we analyze and compare different sharing configurations. Given the importance of directionality in mmWave, and its critical effects on spectrum management, it is vital to accurately characterize the high-dimensional antenna arrays to finely track beamforming and spatial multiplexing gains. Hence, as detailed in \cite{MezzavillaNs3:15}, we use a channel model obtained from a campaign of measurements made in the dense urban environment of New York City \cite{mustafa}. We observed that our detailed model is able to fully capture the potential of spectrum sharing in mmWave, given the fine-grained radiation pattern that is capable of better framing directional gains.

As done in \cite{texas}, and reported in Fig. \ref{shar_conf}, we analyze four different sharing configuration variants.
Prior to describing the different scenarios, we introduce the following sharing definitions.

\textbf{Spectrum Sharing:} Providers share the mmWave bands and thus operate on the same frequencies.

\textbf{Access Sharing:} Users can connect to any other service provider's BS (or a percentage of them).

\textbf{Infrastructure Sharing:} Cell towers are co-located and shared among operators.

Based on these definitions, the sharing paradigm variants are defined as \cite{texas}:

\textbf{Scenario 1:} Our \emph{baseline}, the traditional non-sharing configuration, where operators transmit in their own bands, which are orthogonal to each other, and utilize their own network infrastructures. Users can only connect to the provider they are registered to. These trends are used to evaluate the performance gaps with respect to our sharing scenarios.

\textbf{Scenario 2:} Network providers share the same spectrum, thus simultaneously operating in the same frequencies. We expect to observe an increased performance in terms of per-user rate due to the larger bandwidth available, and a smaller SINR due to the larger number of interferers.

\textbf{Scenario 3:} The BSs of each operator are co-located. Because mmWave signals experience high losses, a dense infrastructure is needed for coverage. Deployment of such infrastructures, separately by each operator, could be too costly, so sharing of base stations and relays would be an efficient way to obtain the required density at lower cost.

\textbf{Scenario 4:} Network operators share the same spectrum. In addition, each provider gives access to any user subscribed to one of the sharing entities. To enable this kind of sharing, high coordination is needed among operators: as shown in Section \ref{subsec:sdn}, SDN and NFV might represent a viable solution to cope with this complexity.

\section{The potential of Resource Sharing}
\label{use}

We compare the four aforementioned scenarios to derive insights and considerations about the impact of resource sharing in mmWave bands. For these numerical results, we consider a system consisting of two networks with identical parameters. Here, each network represents a cellular provider operating in the mmWave frequency band with BS density 30 per km$^2$, which is equivalent to an average cell radius of 103 m \cite{texas}. Each network has a user density of 200 per km$^2$.

The transmit power is assumed to be 30 dBm, with a  noise figure of 7 dB and a carrier frequency of 28 GHz.
The total system bandwidth is 1 GHz. We assume that each network owns a license for 500 MHz.
Recall that in Scenario 1, each network can use only its own spectrum.
In Scenarios 2, 3 and 4, both networks share their spectrum licenses, which results in 1 GHz of available bandwidth to both operators.
BSs and users are deployed in the four scenarios following a Poisson point process (PPP) approach, as done in \cite{ElSawy:13} and \cite{texas}.

\begin{figure}[t!]
    \includegraphics[width=\columnwidth]{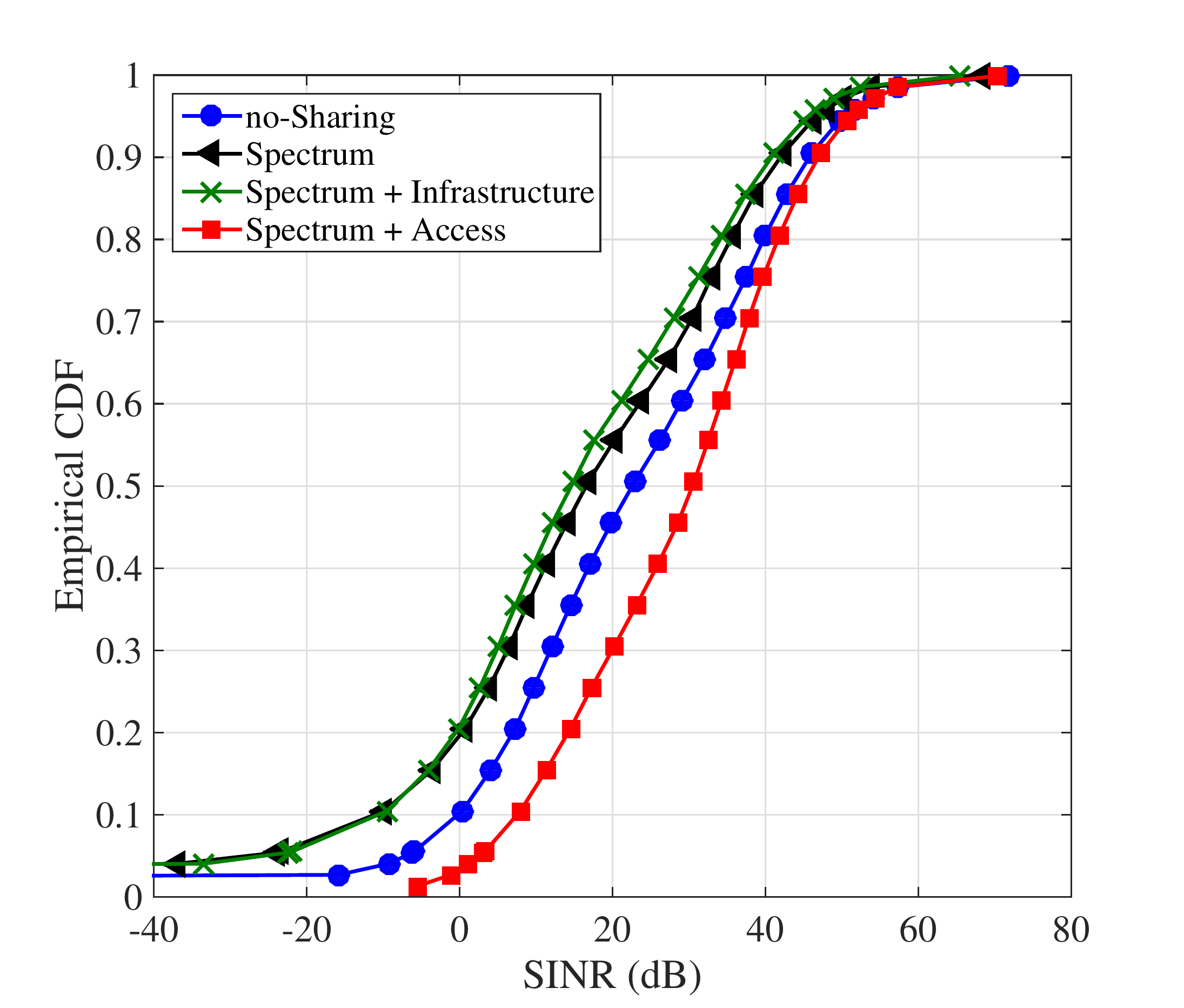}
      \caption{Empirical CDF of the SINR for each sharing configuration.}
      \label{cdf_sinr}
\end{figure}

\begin{figure}[t!]
    \includegraphics[width=\columnwidth]{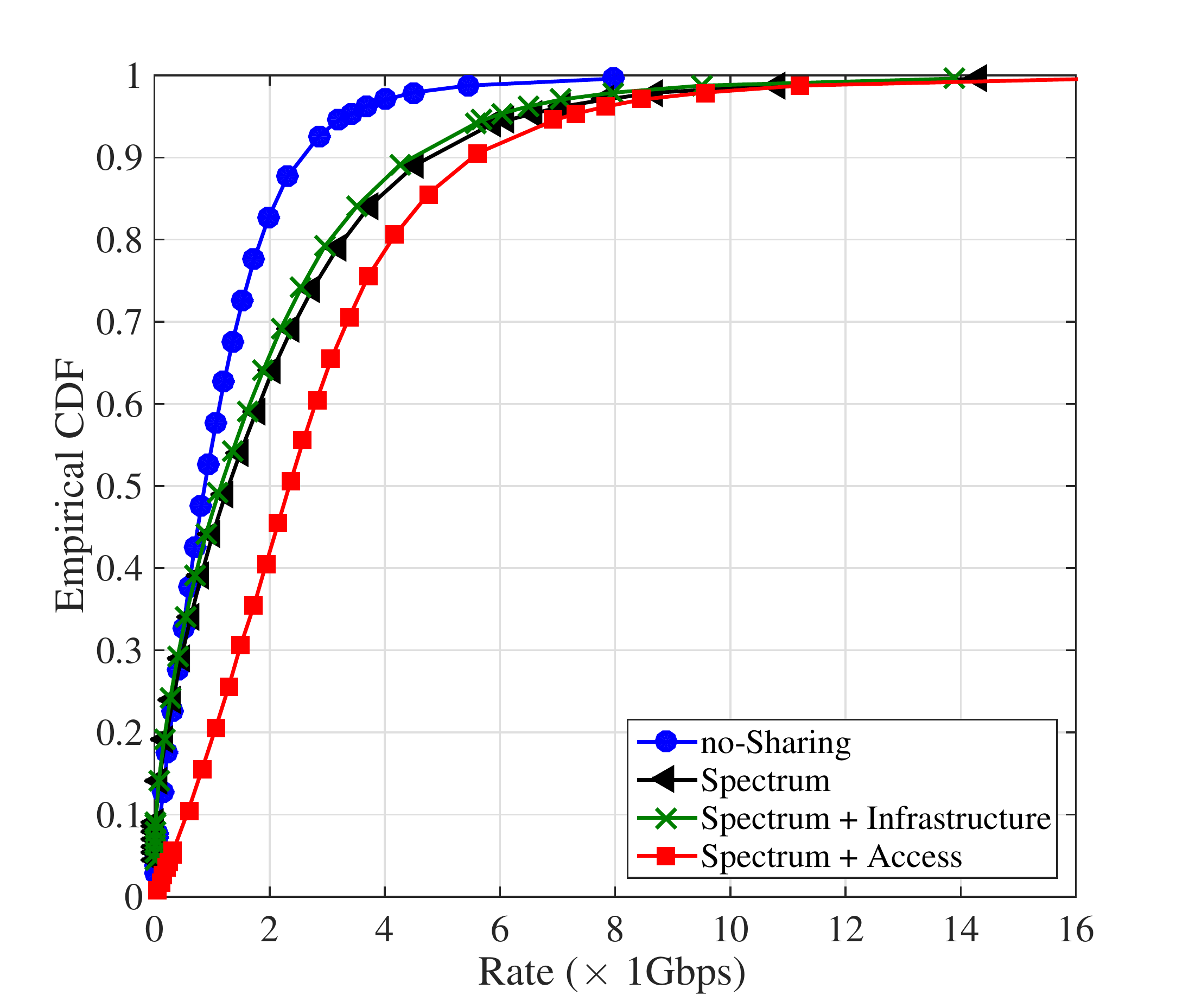}
      \caption{Empirical CDF of the user rate at each sharing configuration.}
      \label{cdf_rate}
\end{figure}

\begin{figure*}[t!]
    \includegraphics[trim={0 5cm 0 0},width=\textwidth]{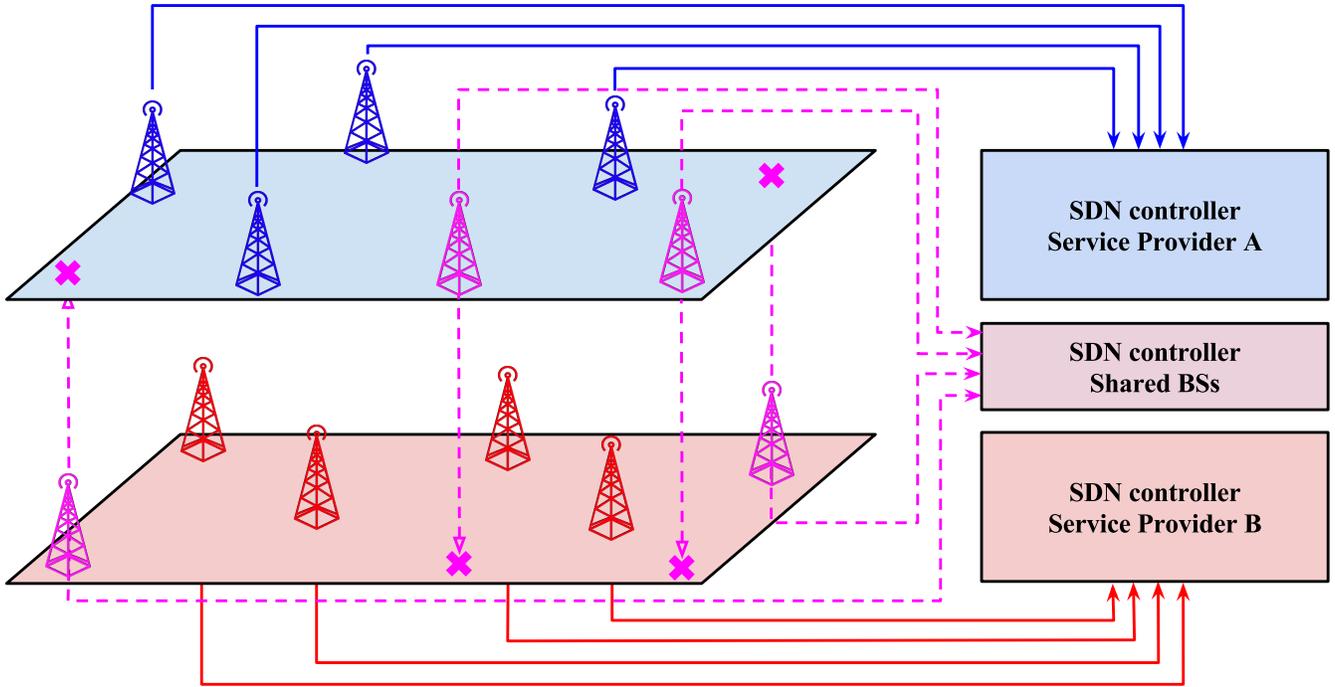}
      \caption{SDN-based architecture to promote access sharing in mmWave cellular netowrks.}
      \label{sdn-ssh}
\end{figure*}

\subsection{Spectrum sharing trends}
Enabling spectrum sharing among operators will, indeed, generate some additional interference. Directional transmissions and increased blockage, which characterize mmWave-based communications, greatly reduce the effects of potential interferers. Nonetheless, allocating the shared spectrum without any type of coordination among operators might result in nearby users being served by proximal base stations over the same frequencies, thus experiencing some level of interference.

In Fig. \ref{cdf_sinr}, we show how this interference-sensitive blind sharing scheme is reflected in terms of decreased signal-to-interference-plus-noise-ratio (SINR). If we compare the empirical cumulative distribution function (CDF) of SINR for the \emph{no-Sharing} and the \emph{Spectrum} scenarios, we observe a slight shift towards reduced received signal quality values. However, given the wider available spectrum, the overall data rate is clearly improved.

The formula used to compute the rate is the following:
\begin{equation*}
R(\gamma)=\eta\alpha (1-\beta) W \textup{log}_2(1+\gamma),
\end{equation*}
where $\eta$ is the Shannon capacity rescaling factor, due to real systems losses (we assume $\eta = 0.5$),  $\alpha$ is the half duplex factor, which is set to $0.5$, $\beta$ is the expected overhead, taken to be $20$ percent of the overall resources. $W$ is the total bandwidth, which is the aggregate spectrum of the sharing providers, and $\gamma$ is the SINR perceived by the user.

\subsection{Reducing deployment costs}
MmWave-aided 5G cellular networks are expected to exhibit high density of BSs to overcome the transmission range limitations in such bands, as discussed in Section \ref{densif}. For this reason, deployment of the network infrastructure, executed independently by each operator, may be very costly, whereas sharing of base stations and relays could be a viable alternative to obtain the required density at reduced cost.

As introduced in Section \ref{potential}, and illustrated in Fig. \ref{shar_conf}, the \emph{Spectrum + Infrastructure} scenario aims at investigating the effects of co-locating different network providers' antennas over the same cell towers. The resulting configuration is like the previous scenario, where operators share their bands, plus an extra feature: each provider deploys its antenna arrays on the same locations as other sharing entities.

By evaluating this type of sharing scenario, we were able to observe the following promising trend. The \emph{Spectrum} and \emph{Spectrum + Infrastructure} results in Figs. \ref{cdf_sinr} and \ref{cdf_rate} exhibit a very similar behavior. In other words, the performance gains observed when providers share their bands is almost identical to the case where operators share also their network deployment costs.
Therefore, it is possible for two operators to share the cell site, thereby reducing their CapEx, with no loss if performance.
A more detailed economical characterization of this approach is left as part of our future work.

\subsection{Inter-operators access coordination}
\label{subsec:sdn}
The key driving motivation behind our last considered sharing scenario, namely \emph{Spectrum + Access}, is \emph{macro-diversity}.

The authors in \cite{macro-div} derived a closed-form expression for the outage probability in mmWave cellular systems, and show how, by exploiting \emph{macro-diversity}, outage decreases with the BS density. However, the deployment cost of cellular systems increases with base station density, which is why we envision a scenario where users registered to \emph{service provider A} can connect to base stations owned and operated by \emph{service provider B}, and vice versa.

As captured in Figs. \ref{cdf_sinr} and \ref{cdf_rate}, this sharing configuration exhibits promising gains, both in terms of coverage and rate. The evaluation of scenarios where operators enable access sharing only to a subset of towers is an interesting area for future work, with the goal of studying smart selection of the \emph{open-access} BSs, aiming at showing that we can achieve solid gains at low sharing percentages.

An SDN-enhanced cellular architecture may represent the right tool to deploy our \emph{Spectrum + Access} scenario. As detailed in \cite{sdn-ssh}, a flexible and programmable network would facilitate reliable and dynamic sharing in 5G systems.
In Fig. \ref{sdn-ssh}, we show how NFV-based core entities would enable transparent inter-operator access for the end users, thus leveraging the potential captured in our simulations. More specifically, we envision a scenario where a dedicated SDN controller manage the BSs that operators agreed to share. Given the flexibility of software-driven network functions, the selection of the shared entities could be dynamical, and thus adapt to user data demand.

For example, as depicted in Fig. \ref{sdn-ssh}, service providers may decide to share $30$ percent of their cell towers with each other. The selection of these entities, which is two per operator in our example, might be driven by the user distribution, which is expected to change in time. Smart selection algorithm design and  numerical insights are left as future work.

\subsection{Interference coordination}
\label{subsec:itf-coo}

\begin{figure}[t!]
    \includegraphics[width=\columnwidth]{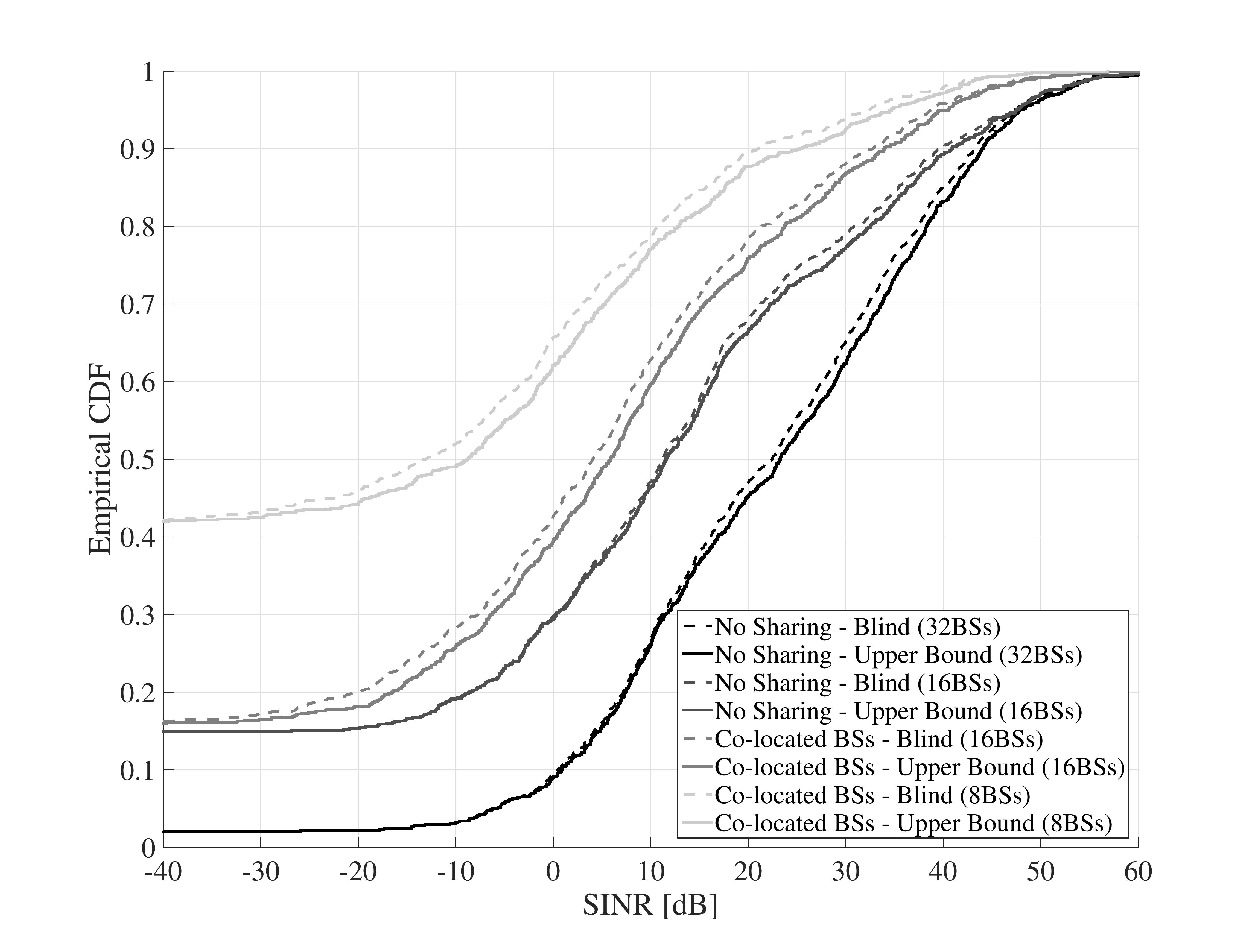}
      \caption{Blind resource allocation Vs. Interference coordination (upper bound).}
      \label{fig:gap}
\end{figure}

Due to the complexity of a coordinated scheme, where operators cooperate to avoid interference, we have adopted a \emph{blind} resource allocation approach, where each BS serves its users fully ignoring the surrounding potential interferers. This is clearly sub-optimal, however, as shown in Fig. \ref{fig:gap}, the performance gap is minimal. This is explained by the fact that interference in the mmWave bands is directional, thus enabling, as described in Section \ref{densif}, a \emph{power}-limited regime where cooperative resource allocation is not needed. In the \emph{upper bound} case, BSs derive an optimal association scheme through a brute force search, which is unrealistic due to the high computational complexity along with the fact that operators cannot know the propagation conditions of all the users. It is interesting to note that our sub-optimal \emph{blind} approach is only slightly outperformed by the \emph{upper bound} even when BSs are co-located, which is the most pessimistic scenario exhibiting the highest directional interference probability. 

\section{Conclusion and Future work}
\label{conclusion}

In this article, we show how resource sharing may represent a killer solution to better leverage the potential of mmWave technology for cellular networks. Huge bandwidth and antenna degrees of freedom offered by these bands can provide statistical multiplexing to accommodate the highly variable nature of 5G traffic. We evaluate the overall performance gains obtained by various sharing configurations, and show how, thanks to a detailed multi-antenna characterization, interference is less dominant, thus providing considerable performance enhancements.
Our key findings show how (i) \emph{spectrum sharing} is a viable solution to better exploit all the potential offered by the mmWave bands, (ii) \emph{infrastructure sharing} can bring almost identical performance gains at reduced deployment cost and, finally, (iii) \emph{access sharing}, which can be enabled by the recent research advances towards adopting a programmable network paradigm through SDN and NFV, is a viable option to leverage \emph{macro-diversity} in mmWave bands and, finally, (iv) we show that we can achieve an overall performance behavior very close to the upper bound through an \emph{uncoordinated} resource allocation scheme.

As future works, we aim at (i) applying practical coordination schemes to inter-operator sharing scenarios, for a more detailed comparison of the benefits of the coordination with the time spent to perform it, (ii) introducing novel business models for resource sharing to better quantify its potential economical impact, and (iii) comparing the overall performance trends obtained at increasing access sharing percentages, in order to propose a model that would be more appealing to service providers.

\bibliographystyle{IEEEtran}
\bibliography{biblio}

\end{document}